\documentclass[preprint,aps,  nofootinbib]{revtex4}
\usepackage{amssymb}
\usepackage{amsmath}
\usepackage{graphicx}

\newcommand\beq{ \begin{eqnarray} }
\newcommand\eeq{ \end{eqnarray} }

\begin{document}

\title{Minimum Shear Viscosity over Entropy Density at Phase Transition?---A
Counterexample}
\author{Jiunn-Wei Chen, Chang-Tse Hsieh, and Han-Hsin Lin}
\affiliation{Department of Physics and Center for Theoretical Sciences, National Taiwan
University, Taipei 10617 }

\begin{abstract}
The ratio $\eta /s$, shear viscosity ($\eta )$ to entropy density ($s)$,
reaches its local minimum at the (second order) phase transition temperature
in a wide class of systems. It was suspected that this behavior might be
universal. However, a counterexample is found in a system of two weakly
self-interacting real scalar fields with one of them condensing at low
temperatures while the other remains in the symmetric phase. There is no
interaction between the two fields. The resulting $\eta /s$ is monotonically
decreasing in temperature despite the phase transition.
\end{abstract}

\maketitle


\section{Introduction}

What are the most perfect fluids in Nature with the smallest shear viscosity
($\eta $) per entropy density ($s$)? Kovtun, Son, and Starinets\ (KSS) \cite%
{KOVT1} suspected that they are a class of strongly interacting conformal
field theories (CFTs) whose $\eta /s=1/(4\pi )$. They even conjectured that $%
1/(4\pi )$ is the minimum bound for $\eta /s$\ for all physical systems.
Ever since the KSS bound was proposed, much progress has been made in
testing this bound and trying to identify the most perfect fluid (see \cite%
{Kapusta:2008vb,Schafer:2009dj} for recent reviews). It is found that $\eta
/s$ can be as small as possible (but still positive) in a carefully
engineered meson system \cite{Cohen:2007qr,Cherman:2007fj}, although the
system is metastable. Also, in strongly interacting CFTs, the universal
value $\eta /s=1/(4\pi )$ is obtained only in the limit of infinite $N$,
with $N$ the size of the gauge group, and infinite t'Hooft coupling limit 
\cite{Son:2007vk}. $1/N$ corrections can be negative, however, \cite%
{Kats:2007mq,Brigante:2007nu}\ and can modify the $\eta /s$ bound slightly 
\cite{Brigante:2008gz,Buchel:2008vz}.

In the real world, the smallest $\eta /s$ known so far belongs to a system
of hot and dense matter thought to be quark gluon plasma just above the
phase transition temperature produced at RHIC \cite{RHIC}\ with $\eta
/s=0.1\pm 0.1(\mathrm{theory})\pm 0.08(\mathrm{experiment})$ \cite%
{Luzum:2008cw}. A robust upper limit $\eta /s<5\times 1/(4\pi )$ was
extracted by another group \cite{Song:2008hj} and a lattice computation of
gluon plasma yields $\eta /s=0.134(33)$ \cite{etas-gluon-lat}. Progress has
been made in cold unitary fermi gases as well. An analysis of the damping of
collective oscillations gives $\eta /s\gtrsim 0.5$ \cite{Schafer,Turlapov}.
Even smaller values of $\eta /s$ are indicated by recent data on the
expansion of rotating clouds \cite{Clancy,Thomas} but more careful analyses
are needed \cite{Schaefer2}.

Previous studies have given some clues about where to find the most perfect
fluid in nature. The first one is to study strongly interacting systems
because strong interaction generally implies small $\eta /s$. The second
clue can be found in a large class of systems where $\eta /s$ goes to a
local minimum near the phase transition temperature ($T_{c}$) \cite%
{Csernai:2006zz,Chen:2006iga,Chen:2007jq}. In particular, $\eta /s$ develops
a cusp(jump) at $T_{c}$ for a second(first) order phase transition and a
smooth local minimum for a cross over. This behavior is seen in QCD with
zero baryon chemical potential \cite{Csernai:2006zz,Chen:2006iga} and near
the nuclear liquid-gas phase transition \cite{Chen:2007xe}. It is also seen
in cold unitary fermi gases \cite{etas-supfluid}, in H$_{2}$O, N, and He and
in all the matters with data available in the NIST database \cite%
{webbook,Csernai:2006zz,Chen:2007xe}. Theoretically, these behaviors can be
reproduced in controlled calculations of weakly interacting real scalar
field theories \cite{Chen:2007jq}. Thus, it was speculated that this feature
is universal. If this is indeed the case, then $\eta /s$ can be used to
probe some parts of the systems which are hard to explore otherwise. For
example, one can try to locate the critical point of QCD by measuring $\eta
/s$ \cite{Lacey:2006bc,Chen:2007xe}.

In this paper, however, we present a counterexample of the $\eta /s$
behavior speculated above. In this model, $\eta /s$ does not go to a local
minimum at the second order phase transition temperature. Our model is a
mixture of two weakly self-interacting real scalar fields with one
condensing at low temperatures while the other remains in the symmetric
phase. There is no interaction between the two fields. The advantage of this
model is that its $\eta /s$ can be computed reliably as in \cite{Chen:2007jq}
because of the small couplings \cite{Jeon}. Other counterexamples have been
asserted previously in literature. One of them is a $\sigma $ model
calculation with a local minimum below $T_{c}$ \cite{Dobado:2009ek}. In this
model, large couplings are used to mimic the case of QCD. Thus, it is not
clear this is due to the failure of the Boltzmann equation at large
couplings \cite{Jeon}, or if the effect is generic. Also, holographic models
have constant $\eta /s$ ($=1/4\pi $) in the limit of infinite $N$ and the
infinite 't Hooft coupling limit. If $1/N$ corrections are added, $\eta /s$
becomes monotonically increasing below $T_{c}$ and a constant above $T_{c}$ 
\cite{Buchel:2010wf}. Our model is a field theory model that one can compute
directly and reliably. Our final result shows that $\eta /s$ does not have
to develop a local minimum at $T_{c}$.

\section{The model}

We will study real scalar theories in cases I-III:%
\begin{eqnarray}
\mathcal{L}_{I} &=&\mathcal{L}_{1},  \notag \\
\mathcal{L}_{II} &=&\mathcal{L}_{2},  \notag \\
\mathcal{L}_{III} &=&\mathcal{L}_{1}+\mathcal{L}_{2},
\end{eqnarray}%
where%
\begin{equation}
\mathcal{L}_{i}=\frac{1}{2}(\partial _{\mu }\phi _{i})^{2}-\frac{1}{2}\mu
_{i}^{2}\phi _{i}^{2}-\frac{1}{4}\lambda \phi _{i}^{4}.
\end{equation}%
The $\eta /s$ of cases I and II are well studied in \cite{Chen:2007jq} and
we follow the treatment there. $\lambda $ and $\mu _{i}^{2}$ are
renormalized quantities and the counterterm Lagrangian is not shown. The
renormalization condition is that the counterterms do not change the
particle mass and the four-point coupling at threshold. We will set $%
0<\lambda \ll 1$ such that the systems are bounded from below and we can
compute to leading order, $\eta $ and $s$, in the\ $\lambda _{i}$
expansions. In case I, $\mu _{1}^{2}>0$ and $\phi _{1}$ stays in the
symmetric phase. The resulting $\eta /s$ is monotonically decreasing in
temperature ($T$) \cite{Chen:2007jq}. In case II, $\mu _{2}^{2}<0$. The $%
\phi _{2}\rightarrow -\phi _{2}$ symmetry is spontaneously broken below the
phase transition temperature $T_{c}$. The resulting $\eta /s$ is
monotonically decreasing when $T<T_{c}$ and becomes\ monotonically
increasing when $T>T_{c}$. Also, $\eta /s$ forms a cusp at $T_{c}$ under the
mean field approximation \cite{Chen:2007jq}.\ 

Because there is no interaction between $\phi _{1}$ and $\phi _{2}$ in case
III, the entropy density is just the sum of the $\phi _{1}$ and $\phi _{2}$
entropy 
\begin{equation}
s_{III}=s_{I}+s_{II}.  \label{s}
\end{equation}%
Analogously, in a linear response theory, the Kubo formula relates $\eta $
to an ensemble average of a correlator 
\begin{equation}
\eta =-\frac{1}{5}\int_{-\infty }^{0}\mathrm{d}t^{\prime }\int_{-\infty
}^{t^{\prime }}\mathrm{d}t\int \mathrm{d}x^{3}\langle \left[
T^{ij}(0),T^{ij}(\mathbf{x},t)\right] \rangle \ ,  \label{Kubo}
\end{equation}%
where $T^{ij}$ is the spacial part of the off-diagonal energy momentum
tensor. $T_{III}^{ij}=T_{I}^{ij}+T_{II}^{ij}$ and $\langle \left[
T_{I}^{ij}(0),T_{II}^{ij}(\mathbf{x},t)\right] \rangle =\left[ \left\langle
T_{I}^{ij}(0)\right\rangle ,\left\langle T_{II}^{ij}(\mathbf{x}%
,t)\right\rangle \right] =0$, such that 
\begin{equation}
\eta _{III}=\eta _{I}+\eta _{II}.  \label{eta_III}
\end{equation}

\begin{figure}[tbp]
\scalebox{0.5}{\includegraphics{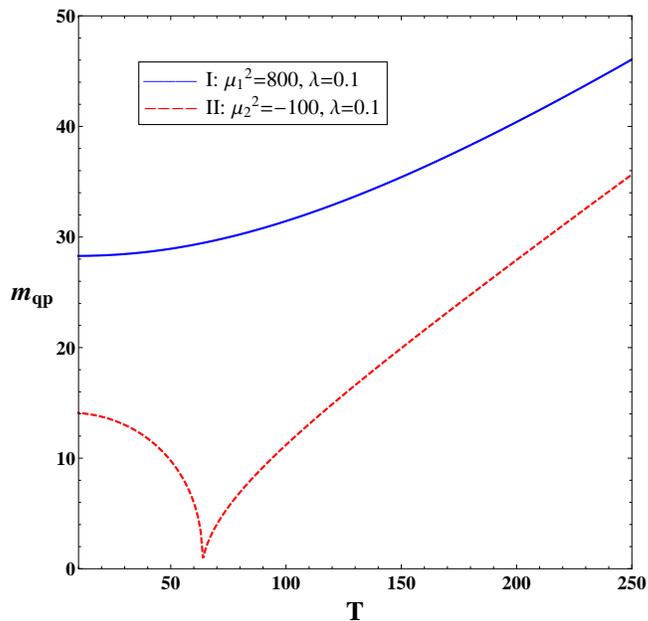}} 
\caption{$m_{qp}$ vs. $T$ for cases \ I (solid curve, without out symmetry
breaking) and II (dashed curve, with symmetry breaking). Parameters can be
in arbitrary units.}
\end{figure}

The high $T$ behavior of $\eta /s$ can be analyzed using the $1/T$ expansion
as in Ref. \cite{Chen:2007xe}. By neglect the slow running of the coupling
constant, the dimensionful quantities $\mu _{i}^{2}$ and $T$ can only
contribute to the dimensionless ratio $\eta /s$ through the $\mu
_{i}^{2}/T^{2}$ combination (note that it is $\mu _{i}^{2}$, not $\mu _{i}$
that appears in the Lagrangian). As $T\rightarrow \infty $, $\eta /s$ has
the following $1/T$ expansion 
\begin{eqnarray}
\frac{\eta _{I}}{s_{I}} &\rightarrow &\frac{c_{1}}{\lambda ^{2}}\left(
1+c_{2}\frac{\mu _{1}^{2}}{T^{2}}+\mathcal{O}(T^{-3})\right) , \\
\frac{\eta _{II}}{s_{II}} &\rightarrow &\frac{c_{1}}{\lambda ^{2}}\left(
1+c_{2}\frac{\mu _{2}^{2}}{T^{2}}+\mathcal{O}(T^{-3})\right) ,
\end{eqnarray}%
where $c_{1}>0$ and $c_{2}>0$. There is no $1/T$ term because as mentioned
above, the result does not depend on the sign of $\mu _{i}$. The $1/T$
expansion of $\eta /s$ in case III has the similar structure 
\begin{equation}
\frac{\eta _{III}}{s_{III}}\rightarrow \frac{c_{1}^{\prime }}{\lambda ^{2}}%
\left( 1+c_{2}^{\prime }\frac{\mu _{1}^{2}+\mu _{2}^{2}}{2T^{2}}+\mathcal{O}%
(T^{-3})\right) .
\end{equation}%
Furthermore, in the limit of $\mu _{2}^{2}=\mu _{1}^{2}$, we have $\eta
_{I}=\eta _{II}$, $s_{I}=s_{II}$, and $\eta _{III}/s_{III}=\eta _{I}/s_{I}$
by Eqs.(\ref{s}) and (\ref{eta_III}). This implies $c_{1}^{\prime }=c_{1}$
and $c_{2}^{\prime }=c_{2}>0$. Therefore, if $\mu _{1}^{2}+\mu _{2}^{2}>0$, $%
\eta _{III}/s_{III}$ is monotonically decreasing in $T$ as $T\rightarrow
\infty $. The question is, whether this behavior persists from large $T$
down to $T_{c}$. Before answering this question numerically, we will try to
understand the behaviors of $\eta $ and $s$ separately.

In Fig. 1, we show the typical $m_{qp,i}$, the effective quasiparticle mass
of case $i$, as a function of $T$. When $T=0$, $m_{qp,I}^{2}=\mu _{1}^{2}$
and $m_{qp,II}^{2}=2\left\vert \mu _{2}^{2}\right\vert $. Then $m_{qp,I}^{2}$
increases for increasing $T$ due to the positive thermal mass effect, while $%
m_{qp,II}^{2}$ decreases to zero at $T_{c}$, and then increases again at
higher $T$. As $T\rightarrow \infty $, $m_{qp,I}^{2}-$ $m_{qp,II}^{2}=\mu
_{1}^{2}-\mu _{2}^{2}>0$. We have chosen the parameters such that $%
m_{qp,I}^{2}>$ $m_{qp,II}^{2}$ at all $T$, which gives $s_{I}<s_{II}$ in
Fig. 2. The cusp in $s_{II}$ is barely visible.

\begin{figure}[tbp]
\scalebox{0.5}{\includegraphics{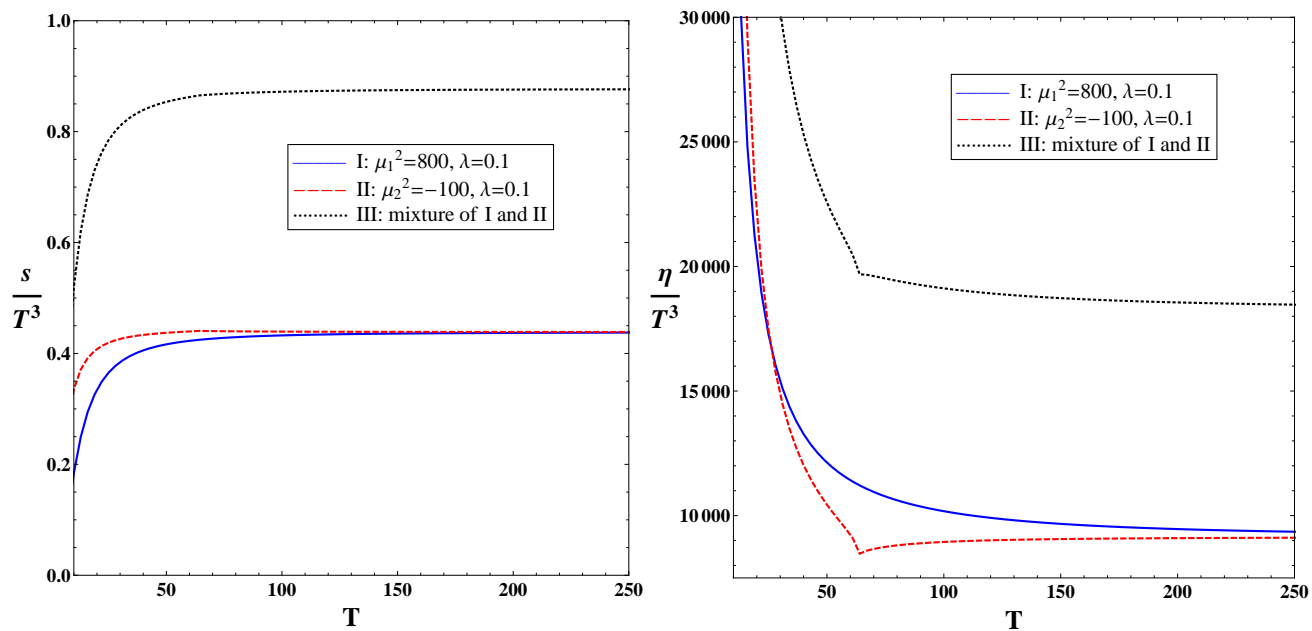}}
\caption{$s/T^{3}$ and $\protect\eta /T^{3}$ for cases \ I (solid curve,
without out symmetry breaking) , II (dashed curve, with symmetry breaking),
and III (dotted curve, the mixture of I and II). Parameters can be in
arbitrary units.}
\end{figure}

In Fig. 2, $\eta /T^{3}$ is also shown. Its behavior is very similar to that
of $m_{qp}$. To further explore this relation, we use the kinetic theory
approximation 
\begin{equation}
\eta \sim \rho vl,
\end{equation}%
where $\rho $, $v$ and $l$ are the quasiparticle density, velocity and mean
free path, respectively. Then using $l\sim 1/nv\sigma $, where $n$ is the
number density and $\sigma $ is the cross section between quasiparticles, we
have 
\begin{equation}
\eta \sim \frac{\rho }{n\sigma }=\frac{\epsilon }{\sigma },
\end{equation}%
where $\epsilon $ is the averaged quasiparticle energy. In a weakly coupled
system, $\epsilon $ can be approximated as a gas of free particles with mass 
$m_{qp}$%
\begin{equation}
\epsilon =\frac{\int d^{3}p\ \epsilon _{qp}f\left( \epsilon _{qp}\right) }{%
\int d^{3}p\ f\left( \epsilon _{qp}\right) }\left[ 1+\mathcal{O}(\lambda )%
\right] ,
\end{equation}%
where $\epsilon _{qp}=\sqrt{p^{2}+m_{qp}^{2}}$ and the Bose-Einstein
distribution $f\left( \epsilon _{qp}\right) =1/\left( e^{\epsilon
_{qp}/T}-1\right) $. In two body collisions, $\sigma $ can be approximated
as 
\begin{equation}
\sigma \sim \frac{\lambda _{eff}^{2}}{\epsilon ^{2}}.
\end{equation}%
Thus, 
\begin{equation}
\eta \sim \frac{\epsilon ^{3}}{\lambda _{eff}^{2}}.  \label{eta}
\end{equation}

\begin{figure}[tbp]
\scalebox{0.5}{\includegraphics{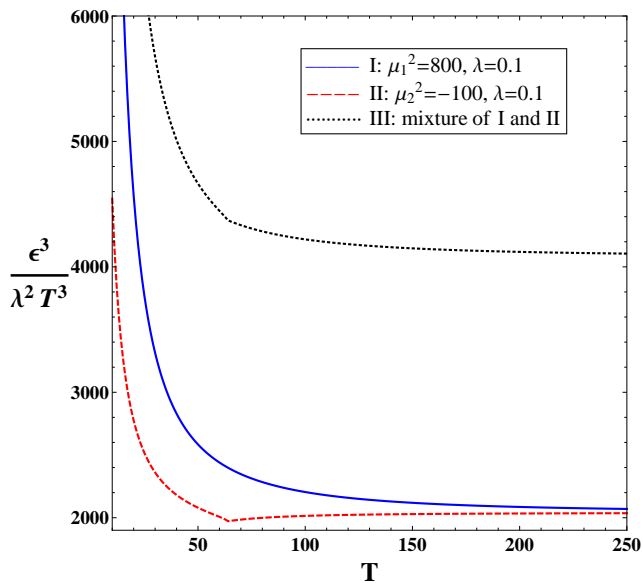}} 
\caption{$\protect\epsilon ^{3}/(\protect\lambda ^{2}T^{3})$ vs. $T$ for
cases I (solid curve, without out symmetry breaking) , II (dashed curve,
with symmetry breaking), and III (dotted curve, the mixture of I and II).
Parameters can be in arbitrary units.}
\end{figure}

The effective coupling $\lambda _{eff}$ is $T$ dependent. The explicit
expression for the scattering amplitude is \cite{Chen:2007xe} 
\begin{equation}
i\mathcal{T}\sim 6\lambda +\left( 6\lambda \left\langle \phi \right\rangle
\right) ^{2}\left[ \frac{1}{s-m_{qp}^{2}}+\frac{1}{t-m_{qp}^{2}}+\frac{1}{%
u-m_{qp}^{2}}\right] .
\end{equation}%
When $T\sim 0$, $s\sim 4m_{qp}^{2}$ and $t\sim u\sim 0$. However, $t\sim
u\sim 0$ causes no momentum redistribution and hence the $t$- and $u$%
-channels have no contribution to $\eta $. Thus, we can approximate $\lambda
_{eff}$ as 
\begin{equation}
\lambda _{eff}\sim \lambda +\frac{6\lambda ^{2}\left\langle \phi
\right\rangle ^{2}}{s-m_{qp}^{2}}.
\end{equation}%
Under this approximation, $\lambda _{eff}$ decreases smoothly from $2\lambda 
$ at $T=0$, to $\lambda $ at $T_{c}$\ and stays constant above $T_{c}$.
Since $\lambda _{eff}$ only varies by a factor $2$, we can further
approximate it by a constant $\lambda $ such that $\eta \sim \epsilon
^{3}/\lambda ^{2}$. As shown in Fig. 3, the $T$ dependence of $\epsilon
^{3}/\lambda ^{2}$ is indeed qualitatively similar to that of $\eta $ in
Fig. 2.

\begin{figure}[tbp]
\scalebox{0.5}{\includegraphics{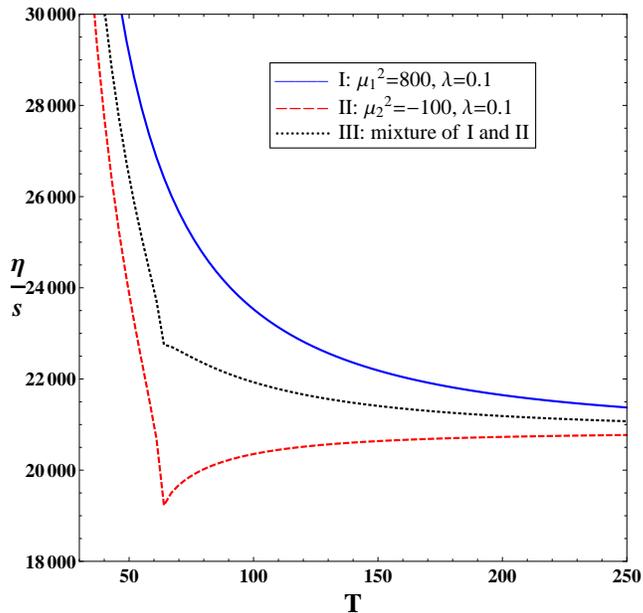}} 
\caption{$\protect\eta /s$ vs. $T$ for cases \ I (solid curve, without out
symmetry breaking) , II (dashed curve, with symmetry breaking), and III
(dotted curve, the mixture of I and II). Parameters can be in arbitrary
units.}
\end{figure}

Finally, we present the $\eta /s$ results in Fig. 4. They are qualitatively
similar to $\eta /T^{3}$. As speculated above, $\eta _{III}/s_{III}$ is
indeed monotonically decreasing both below and above $T_{c}$. This is a
counterexample to the previous speculation that $\eta /s$ goes to a local
minimum at $T_{c}$ of a second order phase transition.

There are some approximations that we have made in this calculation but none
of them should change our conclusion qualitatively. The first one is the
\textquotedblleft Hartree approximation\textquotedblright\ that is used to
neglect all the sunset diagrams below $T_{c}$. This approximation\ is good
when $T\gg \lambda ^{1/2}T_{c}$. At lower $T$, the Hartree approximation is
not reliable. However, as $T\rightarrow 0$, $s$ approaches zero
exponentially (the excitations are massive) while $\eta $ approaches zero
via power laws. As a result, $\eta /s$ is decreasing in cases I-III at low $%
T $. This feature is not affected despite the Hartree approximation used.
The second approximation used is the mean field approximation. Unaccounted
quantum fluctuations can make the result reliable in the region $\left\vert
T-T_{c}\right\vert /T_{c}\lesssim O(\lambda )$. However, this region can be
made arbitrarily small by reducing $\lambda $.

Finally, the end point of a first order phase transition (called a critical
point (CP)) is also a second order phase transition. This is a special kind
of second order phase transition which can be modeled by an effective field
theory whose $\phi _{i}^{2}$ and $\phi _{i}^{4}$ couplings vanishes at $%
T_{c} $, so that the leading coupling is $\phi _{i}^{6}$. Hence the CP case
is different from the cases we consider here. In some systems $\eta $
diverges weakly near a CP.

\section{Conclusion}

The ratio $\eta /s$, the shear viscosity ($\eta )$ to entropy density ($s)$,
reaches its local minimum at the (second order) phase transition temperature
in a wide class of systems. It was suspected that this behavior might be
universal. However, we have presented a counterexample made of a system of
two weakly self-interacting real scalar fields with one of them condensing
at low temperatures while the other remains in the symmetric phase. There is
no interaction between the two fields. The resulting $\eta /s$ is
monotonically decreasing in temperature despite the phase transition.

We thank Brian Smigielski for careful reading of the manuscript. This work
is supported by the NSC and NCTS of Taiwan.



\begin{thebibliography}{99}
\bibitem{KOVT1} P. Kovtun, D.T. Son, and A.O. Starinets, Phys.Rev.Lett. 
\textbf{94},111601 (2005).


\bibitem{Kapusta:2008vb} J.~I.~Kapusta, 
arXiv:0809.3746 [nucl-th]. 


\bibitem{Schafer:2009dj} T.~Schafer and D.~Teaney, 
Rept.\ Prog.\ Phys.\ \textbf{72}, 126001 (2009) [arXiv:0904.3107 [hep-ph]]. 


\bibitem{Cohen:2007qr} T.~D.~Cohen, 
Phys.\ Rev.\ Lett.\ \textbf{99}, 021602 (2007) [arXiv:hep-th/0702136]. 


\bibitem{Cherman:2007fj} A.~Cherman, T.~D.~Cohen and P.~M.~Hohler, 
JHEP \textbf{0802}, 026 (2008) [arXiv:0708.4201 [hep-th]]. 




\bibitem{Son:2007vk} D.~T.~Son and A.~O.~Starinets, 
Ann.\ Rev.\ Nucl.\ Part.\ Sci.\ \textbf{57}, 95 (2007) [arXiv:0704.0240
[hep-th]]. 


\bibitem{Kats:2007mq} Y.~Kats and P.~Petrov, 
JHEP \textbf{0901}, 044 (2009) [arXiv:0712.0743 [hep-th]]. 


\bibitem{Brigante:2007nu} M.~Brigante, H.~Liu, R.~C.~Myers, S.~Shenker and
S.~Yaida, 
Phys.\ Rev.\ D \textbf{77}, 126006 (2008) [arXiv:0712.0805 [hep-th]]. 


\bibitem{Brigante:2008gz} M.~Brigante, H.~Liu, R.~C.~Myers, S.~Shenker and
S.~Yaida, 
Phys.\ Rev.\ Lett.\ \textbf{100}, 191601 (2008) [arXiv:0802.3318 [hep-th]]. 


\bibitem{Buchel:2008vz} A.~Buchel, R.~C.~Myers and A.~Sinha, 
JHEP \textbf{0903}, 084 (2009) [arXiv:0812.2521 [hep-th]]. 

\bibitem{RHIC} I.~Arsene \emph{et al.}, 
Nucl.\ Phys.\ A \textbf{757}, 1 (2005); 
B.~B.~Back \emph{et al.}, 
\emph{ibid.} \textbf{757}, 28 (2005); 
J.~Adams \emph{et al.}, 
\emph{ibid.} \textbf{757}, 102 (2005); K.~Adcox \emph{et al.}, 
\emph{ibid.} \textbf{757}, 184 (2005). 

\bibitem{Luzum:2008cw} M.~Luzum and P.~Romatschke, 
Phys.\ Rev.\ C \textbf{78}, 034915 (2008) [arXiv:0804.4015 [nucl-th]]. 

\bibitem{Song:2008hj} H.~Song and U.~W.~Heinz, 
J.\ Phys.\ G \textbf{36}, 064033 (2009) [arXiv:0812.4274 [nucl-th]]. 

\bibitem{etas-gluon-lat} 
H.~B.~Meyer, 
Phys.\ Rev.\ D \textbf{76}, 101701 (2007), arXiv:0704.1801 [hep-lat]. 


\bibitem{Schafer} T. Schafer, Phys. Rev. A 76, 063618 (2007).

\bibitem{Turlapov} A. Turlapov, J. Kinast, B. Clancy, L. Luo, J. Joseph, and
J. E. Thomas, J. Low Temp. Phys. 150, 567 (2008).

\bibitem{Clancy} B. Clancy, L. Luo, J. E. Thomas Phys. Rev. Lett. 99 140401
(2007) [arXiv:0705.2782 [condmat.other]].

\bibitem{Thomas} \qquad J.E. Thomas, Nucl. Phys. A 830, 665c (2009).

\bibitem{Schaefer2} T. Schaefer, C. Chafin, e-Print: arXiv:0912.4236
[cond-mat.quant-gas]; T. Schaefer, e-Print: arXiv:1008.3876
[cond-mat.quant-gas].

\bibitem{Csernai:2006zz} L.~P.~Csernai, J.~I.~Kapusta and L.~D.~McLerran, 
Phys.\ Rev.\ Lett.\ \textbf{97}, 152303 (2006).

\bibitem{Chen:2006iga} J.~W.~Chen and E.~Nakano, 
Phys.\ Lett.\ B \textbf{647}, 371 (2007). 

\bibitem{Chen:2007xe} J.~W.~Chen, Y.~H.~Li, Y.~F.~Liu and E.~Nakano, Phys.
Rev. D76, 114011(2007).%

\bibitem{etas-supfluid} 
T.~Schafer, 
arXiv:cond-mat/0701251; 
G.~Rupak and T.~Schafer, 
arXiv:0707.1520 [cond-mat.other]. 

\bibitem{webbook} E.W.~Lemmon \emph{et al.}, Thermophysical Properties of
Fluid Systems, in \textit{NIST Chemistry WebBook}, NIST Standard Reference
Database Number 69, Eds. Linstrom P.G. \& Mallard, W.G., March 2003
(http://webbook.nist.gov).

\bibitem{Chen:2007jq} J.~W.~Chen, M.~Huang, Y.~H.~Li, E.~Nakano and
D.~L.~Yang, 
Phys.\ Lett.\ B \textbf{670}, 18 (2008) [arXiv:0709.3434 [hep-ph]]. 

\bibitem{Lacey:2006bc} R.~A.~Lacey \textit{et al.}, 
Phys.\ Rev.\ Lett.\ \textbf{98}, 092301 (2007); arXiv:0708.3512. 



\bibitem{Jeon} S. Jeon, Phys. Rev. D \textbf{52}, 3591 (1995); S. Jeon and
L. Yaffe, Phys. Rev. D \textbf{53}, 5799 (1996).

\bibitem{Dobado:2009ek} A.~Dobado, F.~J.~Llanes-Estrada and
J.~M.~Torres-Rincon, 
Phys.\ Rev.\ D \textbf{80}, 114015 (2009) [arXiv:0907.5483 [hep-ph]]. 


\bibitem{Buchel:2010wf} A.~Buchel and S.~Cremonini, 
arXiv:1007.2963 [hep-th]. 
\end{thebibliography}
\end{document}